\documentclass[preprint]{aastex62}
\graphicspath{{./}{figures/}}

\received{-}
\revised{-}
\accepted{-}

\submitjournal{ApJS}

\shorttitle{Non-Linear Solar-like Stars}
\shortauthors{Tapia \& DelaLuz}
\begin{document}

%\title{Non-Linear Convergence of Solar-like Stars Atmospheres using semiempirical Models of the Solar Chromosphere}
\title{Non-Linear Convergence of Solar-like Stars Chromospheres using Millimeter, Sub-millimeter, and Infrared Observations}

\correspondingauthor{Victor De la Luz}
\email{vdelaluz@enesmorelia.unam.mx}

\author[0000-0001-9132-7196]{F. Tapia-V\'azquez}
%\affil{Escuela Nacional de Estudios Superiores Unidad Morelia, Universidad Nacional Aut\'onoma de M\'exico, Mexico.}
\affiliation{Instituto de Radioastronom\'\i a y Astrof\'\i sica, Universidad Nacional Aut\'onoma de M\'exico, \\P.O. Box 3-72, 58090, Morelia, Michoac\'an, M\'exico.}

\author{V. De la Luz}
%\affil{Escuela Nacional de Estudios Superiores Unidad Morelia, Universidad Nacional Aut\'onoma de M\'exico, Mexico.}
\affiliation{Escuela Nacional de Estudios Superiores Unidad Morelia, 
Universidad Nacional Aut\'onoma de M\'exico, 
Morelia, 58190, M\'exico}

\begin{abstract}

In this work, we present a new methodology to fit the observed and synthetic spectrum of solar-like stars at millimeter, submillimeter and infrared wavelengths through semiempirical models of the solar chromosphere. 
%We are using Levenberg-Marquardt as a Non-Linear method, 
We use the Levenberg-Marquardt algorithm as a Non-Linear method, PakalMPI as the semiempirical model of the solar chromosphere, and recent observations from the Atacama Large Millimeter/submillimeter Array (ALMA) of Alpha Centauri A as a test case. Our results show that we can use solar chromospheric semiempirical models as an input model to reproduce the observed spectrum of solar-like stars. The new profiles show similarities to the solar chromosphere as a minimum of temperature (without the restriction from CO emission) and a plateau in the high chromosphere. Our method provides a new fast numerical tool to estimate the physical conditions of solar-like stars. 

\end{abstract}

%\keywords{editorials, notices --- 
%miscellaneous --- catalogs --- surveys}


\section{Introduction} \label{sec:intro}

The particular characteristics of the emission at millimeter, submillimeter, and infrared wavelengths in the solar chromosphere allow us to estimate their temperature and density using indirect methodologies like semiempirical models \citep{vernazza81, fontenla93, avrett}. These models are an important tool for a wide set of studies e.g. solar and stellar chromospheres \citep{loukitcheva, linsky17}, temperature minimum \citep{liseau13, delaluz14}, solar flares \citep{machado80, trottet} and Sun component features \citep{fontenla06}.

In terms of solar atmospheric temperatures, the semiempirical models predict values close to photospheric temperatures that decreases until a minimum, then increase dramatically until the coronae \citep{vernazza81,avrett} in a thin layer of around $2200$ km. This layer is known as the solar chromosphere. The chromosphere remains observable by different spectral ranges that include Ultraviolet (UV) in the continuum and line emission, in the visible (mainly H$\alpha$) and in the millimeter, submillimeter and infrared wavelengths. This last wavelength range becomes more important as improvements in the sensitivity and space resolution in modern radio telescopes \citep{nakajima, kudaka, wedemeyer16}.

The process to compute an atmospheric model begin testing the initial conditions of a semiempirical model \citep{vernazza76, carlsson92, dere97, delaluz10}. The process includes three steps: i) small modifications in an initial temperature profile, ii) compute the density and pressure required to guarantee hydrostatic, hydrodynamic or magnetohydrodynamic equilibrium, and iii) compute the synthetic spectrum to compare with observations. Differences between synthetic and observed spectrum are solved making changes in the atmospheric temperature profile at different altitudes until synthetic and observed spectrum converges.   
The number of iterations to converge both observations and the synthetic spectrum restrict the precision of the computation.

Recent observations at sub-mm wavelengths of Solar-like stars allow computing for the first time the %temperature profile 
stellar atmosphere at chromosphere altitudes using as start model the solar chromospheric by millimeter-infrared wavelength range observations \citep{liseau16}. 
However, the differences between the solar chromosphere and the structure of the atmosphere of Solar-like stars requires thousands of iterations to fit observations and the synthetic spectrum. 

In this work, we present a new non-linear model to converge automatically observed and synthetic spectrum of solar-like chromospheres at millimeter-infrared wavelength. We used the Levenberg-Marquardt (LM) method to fitting observations and synthetic spectrum using PakalMPI as parametric function \citep{delaluz11}. The LM algorithm use three parameters: a set of points in the electromagnetic spectrum as the independent variable ($\nu$) in the LM scope, PAKALMPI as function model ($f(\nu,T_r)$), and the temperature model ($T_r$) as a set of discrete points used to computed 
\begin{equation}\label{eq1}
T_b=f(\nu,T_r), 
\end{equation}
where $T_b$ is the synthetic spectrum.
In our case, the observed spectrum that corresponds with the frequencies is the dependent data ($T^o_b$).
Our model does not take into account limb brightening contribution. We are computing the brightness temperature in the center of the stellar disk.

In Section \ref{sec:independent} we present the region of the spectrum under study (independent variable). In Section \ref{sec:parametric} present our parametric semiempirical model PakalMPI (function model). In Section \ref{sec:dependent} the set of observations of Solar-like stars used as test cases (dependent data). In Section \ref{sec:LM} we describe the general approximation and restriction of the LM method used in this work. In Section \ref{sec:test} we show the results of the case test. Finally in Section \ref{sec:discuss} we present our discussion and conclusions.

\section{The independent variable: The millimeter,sub\-millimeter, and infrared wavelengths}\label{sec:independent}
The independent variable in the LM scope represents the set of data that are evaluated in the function model. In this work, the independent variable ($\nu$) is a set of frequencies of the electromagnetic spectrum at millimeter, submillimeter, and infrared wavelengths
\begin{equation}
    \nu = \{ \nu_1, \nu_2,...,\nu_k\},
\end{equation}
where $k$ is the number of points in the observed electromagnetic spectrum and $\nu_i$ are central frequencies of radio observations constrained to cover the 10 bands by ALMA (from $35$GHz to $950$GHz) and Spitzer ($12.5$THz):
$$ 35\mbox{GHz} \leq \nu_i \leq 12.5\mbox{THz}.$$
This range of frequencies allow us to use the tomographic properties of the emission to estimate the radial temperature and density profile evolved in the synthetic electromagnetic spectrum \citep{delaluz16}.

\section{The parametric function model: PakalMPI}\label{sec:parametric}
LM requires a function model that receives the independent variable ($\nu$) and a set of parameters ($T_r$) that fit the model (Equation \ref{eq1}). In our case, we modify PakalMPI model \citep{delaluz10,delaluz11} to work as parametric function. PakalMPI model uses Bremsstrahlung \citep{dulk85, gudel2002}, Inverse Bremsstrahlung \citep{golovinskii1980}, and H- \citep{zheleznyakov96} mechanisms as source of opacity. However, Bremsstrahlung and H- are the most important mechanism of thermal emission in the chromosphere \citep{delaluz11}.

The set of parameters are temperature points that represent the radial temperature profile under study
$$T_r = \{T_1,T_2,..,T_n\},$$
where $n$ is the number of altitudes in the radial temperature profile and  $T_i$ are the ordered radial temperatures against distance over the photosphere from a semiempirical model.
We used C7 model \citep{avrett} as initial condition (radial temperature and density) because is a consistent model of the solar chromosphere especially on the temperature minimum \citep{delaluz14}. The C7 model is a one-dimensional and time independent model of the average quiet-Sun. This model replaced the earlier model C of \citet{vernazza81} and take into account observations of the Extreme Ultraviolet (EUV) spectrum and two measurements in the submillimeter region to get a good average of the temperature minimum. However, this model does not reproduce correctly the observations in the submillimeter, millimeter and radio range \citep{delaluz16}. \citet{delaluz14} showed that by modifying the temperature profile it is possible compute a better fit in this region of the spectrum. The new model obtained was compatible with the theory presented by \citet{avrett}. 

The temperature profile is passed as arguments together with the frequency ranges to PakalMPI. Then, PakalMPI compute in three steps the synthetic spectrum: i) compute a new radial density profile iteratively using the radial temperature profile to satisfy the hydrostatic equilibrium,
%compute the hydrostatic equilibrium using the input temperature model {\bf in this process the radial density is modified to satisfy the hydrostatic equilibrium}, 
ii) compute the ionization stages of a set of atoms that represents the chemical composition of the atmosphere, and iii) solves the radiative transfer equation in the set of requested frequencies. The two last steps can be computed in parallel using the Message Passing Interface (MPI) methodology. Detailed information about the steps can be found in \citet{delaluz10}.

PakalMPI output includes the synthetic spectrum, the radial profiles of all the ions, temperature, density, and pressure of the atmosphere in equilibrium. This output is used in further steps of our methodology.

\section{The dependent data: Observations of Solar-like stars at millimeter, submillimeter, and infrared wavelengths}\label{sec:dependent}
In this work, the dependent variable is represented by the observed spectrum
\begin{equation}
    T_b = \{ T_b ^{\nu_1}, T_b ^{\nu_2},...,T_b ^{\nu_k}\},
\end{equation}
where $T_b^{\nu_k}$ is a observed point in the electromagnetic spectrum.

The development of new infrastructure (Atacama Large Millimeter/submillimeter Array (ALMA), Large Millimeter Telescope (LMT), etc.) allow us to study main sequence stars in mm - sub-mm regime.
The emission at these wavelengths is originated in the chromosphere mainly to free-free emission \citep{dulk85, loukitcheva, wedemeyer16}, and neutral interaction \citep{zheleznyakov96} but in general, it is not well-constrained \citep{cranmer}.
The first binary stellar system that has been observed and studied in this regime has been Alpha Centauri ($\alpha$ Cen) located at $1.3384\pm0.0011$pc \citep{kervella17}. The system $\alpha$ Cen is composed of two stars: $\alpha$ Cen A (G2 V) and $\alpha$ Cen B (K1 V). 
In both, the minimum temperature was observed by \citet{liseau13} in the far-infrared. \citet{liseau16} showed that an atmospheric structure similar to the Sun can be adapted to solar-like stars.

In order to test our model, we have chosen $\alpha$ Cen A since it is often considered a solar twin \citep{cayrel96} and has observations from the far infrared to the millimeter \citep{liseau16}. 

\section{Levenberg-Marquardt approach: The Kinich-Pakal model}\label{sec:LM}
There are several implementations of the Levenberg-Marquadt algorithm \citep{marquardt, press86, pujol07}. In this work we use least square fit (LsqFit) package developed by Julia Optim team \citep{mogensen18}.

The model that we present in this work is called Kinich-Pakal (KP). The aim of KP is to compute the best model of stellar atmosphere to fit the synthetic spectrum to observations at millimeter, submillimeter, and infrared wavelengths.

\begin{figure}[ht!]
\plotone{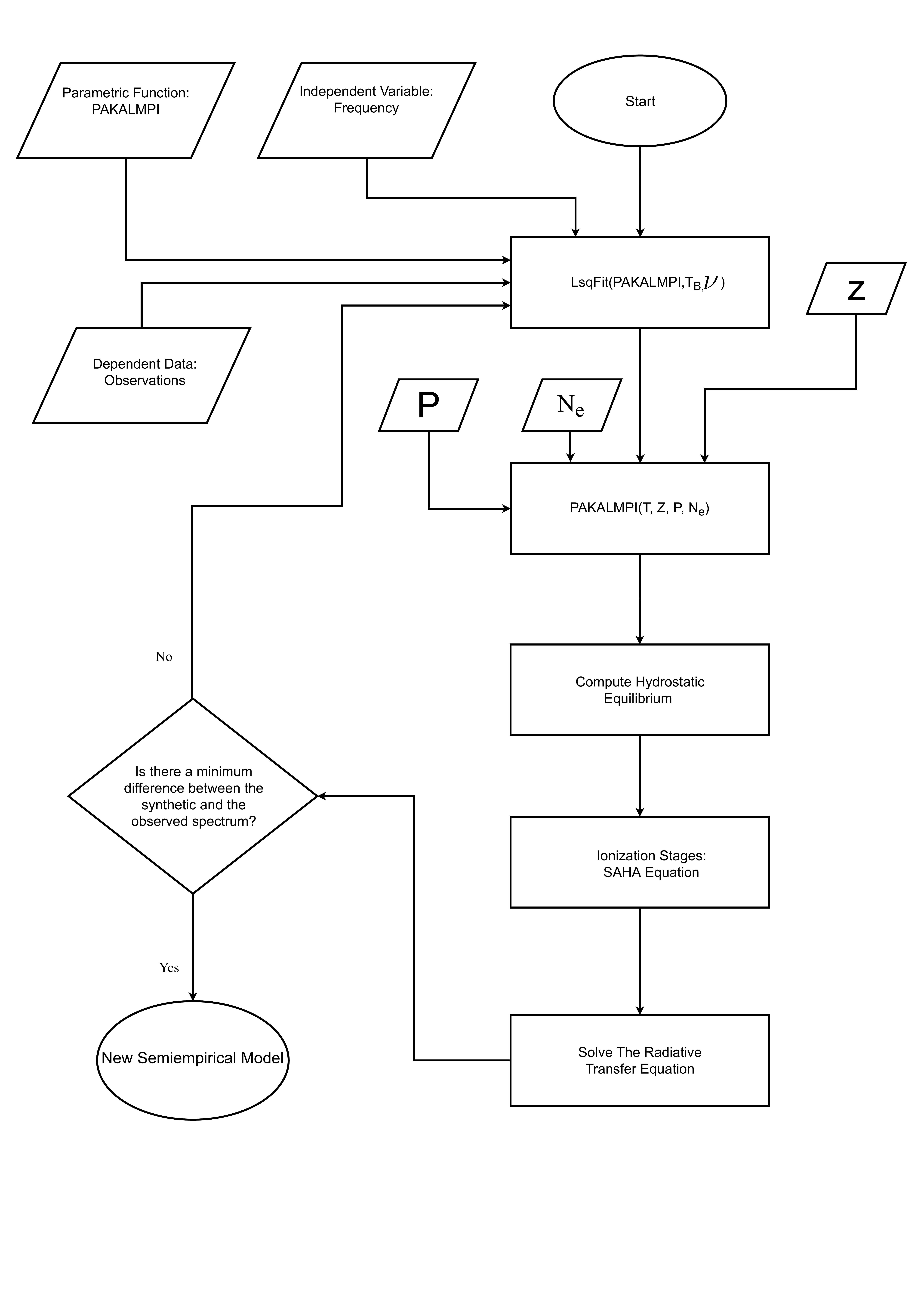}
\caption{Diagram of operation for KP. KP uses a parametric function (PAKALMPI), a set of dependent data (observations) and independent variables (frequency) as inputs. LsqFit function starts the Levenberg-Marquart algorithm. PAKALMPI compute the hydrostatic equilibrium and solve the radiative transfer equation. LM receives the computed spectrum from PAKALMPI and compares it with observations. If the error is higher than predefined $\delta$ then LM modifies the temperature profile and repeats the process. If the error is less than $\delta$ then KP produces a new semiempirical model. \label{fig:kpdiagram.pdf}}
\end{figure}

In Figure \ref{fig:kpdiagram.pdf} we show the pseudo code of KP. We start reading the initial conditions of a single model that includes: radial temperature, Hydrogen density, and pressure. KP read the radial profiles and the observations and start the computations iterative using PakalMPI as a function model. 

In each iteration, KP computes the hydrostatic equilibrium of the input atmosphere and compute their synthetic spectrum. The computations of the synthetic spectrum is performed in parallel for each frequency.
If the differences between the observed and the synthetic spectrum are greater than an epsilon stop flag, then the computed atmosphere becomes the initial condition for the next iteration and the process continues to step forward. 

KP has several flags that constrain the Levenberg-Marquardt method. The constraints include: set the global limits of the solution, limit the solution in a defined range, and the absolute error tolerance against the set of observations. The performance of KP depends of the flag constraints and is studied in the following section.

\section{Test Case: $\alpha$ Centauri A}\label{sec:test}
$\alpha$ Cen A and B was the first stellar system resolved in submillimeter by ALMA \citep{liseau15}. Table \ref{tab:almaobs} show the properties of the observations of $\alpha$ Cen A during the Cycle $2$ of ALMA between July $2014$ and May $2015$.  In Table \ref{tab:almaobs2} we show additional observations of $\alpha$ Cen A. 
In the following subsections, we describe the set of test used by KP.

\subsection{Infrastructure for the computations}
For the following tests, we run KP in a single node of the cluster of the Center of Supercomputing of Space Weather of the National Laboratory of Space Weather in Mexico (LANCE). The node of the cluster have $40$ threads with Intel Xeon E52670 v2 a $2.5$GHz, $128$GB of RAM (DDR3-$1866$) and $4 TB$ for storage. The operating system is Linux  $4.2.0$ SMP for $64$bits.

\subsection{Test of auto convergence}\label{sub:autoconvergence}
Before starting these simulations with the real observed spectrum, we test the model using a synthetic spectrum produced by a solar atmosphere in hydrostatic equilibrium.
%The Figure \ref{fig:tinputoutput.pdf} show the first test of KP focused in the auto convergence. 

We use as input atmospheric model a set of radial profiles with variations of a pre-calculated atmosphere in equilibrium over the photosphere (z). We use the C7 model with variation of temperature of $-1000$K, $-100$K, $-10$K, $+10$K, $+100$K, and $+1000$K as shown in Figure \ref{fig:tinputoutput.pdf}a. We replace the observed spectrum used in this test with a synthetic spectrum computed with PakalMPI for the C7 model (Test Spectrum). The idea of replace the observed spectrum is to keep under control the input and the output of the model.

KP computed successfully these models in around $3$ hour using a single node with $28$ threads (one for each frequency). Table \ref{tab:ss} show the frequencies and the computed synthetic spectrum used in this test.

%We found that changes in the radial temperature profile between {\bf $-1000$K} and $+100$K are minimal (Figure \ref{fig:tinputoutput.pdf}b). {\bf For the case of $+1000$K we found that the original $T_r$ is not completely recovered, i.e. our methodology presents an upper boundary of $+1000$K for the initial condition.}

In order to determinate the region where the emission is generated, we include in Figure \ref{fig:tinputoutput.pdf}c the normalized Contribution Function (CF) defined by
\begin{equation}
    \mbox{CF}= j_{\nu}\exp{(-\tau_{\nu}),}
\end{equation}
where $j_{\nu} =  \kappa_{\nu} B(T)$ and
\begin{equation}
\tau_{\nu} =  \int\kappa_{\nu}dz,
\end{equation}
is the optical depth. We found that our model computes the emission between $140$km and $2120$km for $12.5$THz and $35 $GHz respectively. The gray regions in Figure \ref{fig:tinputoutput.pdf}a-c shows the outer boundary of the model where the convergence method can not be applied directly.
On the other hand, the synthetic spectrum computed with these models fit the Test Spectrum (Figure \ref{fig:tinputoutput.pdf}d). The only result that not converge completely to the initial condition is the $+1000$K model (segmented black line in Figure \ref{fig:tinputoutput.pdf}b). The auto convergence test show that our methodology has an upper boundary for the initial condition of $+1000$K.

In the following tests, we use the real spectrum of $\alpha$ Cen A with different restrictions in KP to know the stability of the model.

\subsection{Test without Restrictions (KP2019A)}
In the second test, we use real observations of $\alpha$ Cen A using C7 as an input model without restrictions (i.e. the temperature model can take any value, even negative). KP makes $12730$ iterations in $49.5$ hours in order to obtain the model KP2019A. The results are shown in Figure \ref{fig:kp.pdf}a-b as dotted line. KP fixes the observed spectrum (Figure \ref{fig:kp.pdf}a) but the atmospheric model presents changes in the radial temperature that are very complicated to explain physically. The model shows the following characteristics: high temperature at around $1500$km over the photosphere $T_r>11500$K, negative temperatures at $1950$km, and abrupt changes of temperature that reaches values between $3\times10^5$K and $-2\times10^5$K that are physically impossible.
Analyzing in detail the ray path of the solution of the radiative transfer equation, we found that negative temperatures are due Bremsstrahlung opacity. This opacity has the following form:
$$\kappa_\nu = C \frac{n_en_i}{T^{3/2}}\nu^2,$$
where $C$ is a constant, $n_e$ is the electron density, $n_i$ the ion density, $T$ the temperature, and $\nu$ the frequency. For temperatures less than $3000$K the ion density of all the species that we study drops almost to zero ($n_i\approx0$) due there is not enough energy to ionize the medium, especially Hydrogen. That's means that the atmosphere becomes optically thin. In the other way, when the temperatures are higher than $2\times10^4$K all the medium is practically ionized, then the opacity can be rewritten as
$$\kappa_\nu \approx C \frac{n_e^2}{T^{3/2}}\nu^2,$$
the $n_e$ reaches a limit, however, the temperature can increase and as a consequence, the opacity again drops to zero.
In both cases: low and high temperatures the atmosphere becomes optically thin. These results show that exist solutions computed by KP in the radiative transfer equations without physical interpretations. In the following test, we constrain the minimum temperature in the solution to try to solve this feature of the model.

\subsection{Test Constraining Positive Temperatures (KP2019B)}
In this test, we constrain the lower temperature limit in $3000$K. KP produced the KP2019B model after computed $11300$ models in around $44$ hours. In the Figure \ref{fig:kp.pdf}a-b we show the results as dashed-point line. We observe that the computed spectrum not exactly fix the observations of $\alpha$ Cen A (Figure \ref{fig:kp.pdf}a) like the last test. The radial temperature profile again shows temperatures without physical interpretations (Figure \ref{fig:kp.pdf}b). After $1600$km, the radial temperature increase suddenly with temperatures around $10^5$K. 
%After this second plateau the temperature drops to two temperature minimums in less than 50 km. Finally the model of temperature reaches coronal temperatures.

\subsection{Test with Bounded Temperature (KP2019C)}
We bounded the temperature taking both: the initial model as reference and the results of the Subsection \ref{sub:autoconvergence}. We use C7 model and a boundary of $\pm1000$K as constraining. KP computed $7968$ models in around $31$ hours to produce our model called KP2019C. In Figure \ref{fig:kp.pdf}a-b we show the results as dashed line. Figure \ref{fig:kp.pdf}a shows that the spectrum computed by KP is close to fit the observed spectrum. However, we can observe that for low frequencies the model not fix exactly the observed spectrum. The temperature model (Figure \ref{fig:kp.pdf}b) shows similar C7 model shape  but with three main differences: i) at photospheric altitudes themKP2019C show a higher temperature than C7 model, ii) the temperature minimum of KP2019C is closer to the photosphere and shows a lower temperature of around $3200$K than C7 model, and iii) the temperature in the high chromosphere of KP2019C (after $1500$km) is higher than C7 by $1000$K.
In the following test, we analyze the impact of the spatial resolution of the radial temperature profile.

\subsection{Testing the Spatial Resolution (KP2019D)}
The C7 model starts with $140$ layers divided into three parts: photosphere ($20$ layers), chromosphere ($81$ layers), and corona ($39$ layers) but in this test, we interpolate the model to increase the density 4 times in the chromospheric region up to $324$ layers. KP2019D model is composed of $383$ layers. 

We used the same constraints as in the last subsection. To calculate the final spectrum, KP computes $23585$ models in around $156$ hours. The cyan solid lines in Figure \ref{fig:kp.pdf}a-b shows the results of KP2019D model. In figure \ref{fig:kp.pdf}a, we plot the spectrum obtained by the model showed in Figure \ref{fig:kp.pdf}b. We observe the same similarities of KP2019C but with changes in temperatures with more detailed spatial resolution. For example, the peak of temperature of KP2019C at around $1000$km now is a set of spikes. 

\subsection{Smoothing Model (KP2019E)}
We smooth the radial temperature profile from KP2019D using the filter of Savitzky-Golay with a nine degrees polynomial \citep{savitzky} to know the effect of these spikes in the spectrum. This last profile was equilibrated again with PakalMPI to guarantee the hydrostatic equilibrium. The results are plotted in red colors in the same Figure \ref{fig:kp.pdf}a-b. We called this model KP2019E. 

In the model KP2019E, we eliminate the spikes and compute their synthetic spectrum using PakalMPI. In Figure \ref{fig:kp.pdf}a we observe the results of eliminating the spikes in the temperature model. At low frequencies, the spectrum has a better fix. At frequencies lower than $300$GHz the spectrum not fix exactly the observations but are inside of the observations error limits for all the points of the spectrum, except at $700$GHz that correspond to the Band 9 of ALMA. This observation is the responsible of the increase of temperature in the radial profile between $450$km and $1100$km over the photosphere.  The model KP2019E shows a continue profile without spikes. The temperature minimum is lower ($\sim3000$K) and closer to the photosphere ($\sim500$km) than C7 model. The radial temperature of KP2019E after $1500$km is higher than C7 model. 

In Figure \ref{fig:kp.pdf} we show the results of computing the CF (panel c) and optical depth (panel d) for the KP2019E model at $35$GHz, $97.5$GHz, $233$GHz, $679$GHz, and $12491.5$GHz. We found that for the case of $12491$GHz the CF have a maximum at altitudes closed to $250$km
over the photosphere. For the other cases, the CF shows three peaks at $679$GHz, two peaks at $233$GHz, 
and one peak at $97.5$GHz and $35$GHz. These peaks in the CF shows similarities with the reported solar CF \citep{selhorst19} that show two peaks (around $230$GHz) and one peak (around $100$GHz).

\section{Discussion and Conclusions}\label{sec:discuss}
In this work, we present KP a non-linear method to compute a stellar atmosphere to fix observed and synthetic spectrum trough a semiempirical model of the chromosphere at millimeter, submillimeter and infrared wavelengths. KP uses the Levenberg-Marquardt algorithm to minimize the differences between synthetic and observed spectrum running PakalMPI to equilibrate hydrostatically the atmosphere and compute their synthetic spectrum. KP uses MPI libraries, C, and Julia Languages to solve the model in parallel where it is possible to divide the computations.  

\begin{figure}[ht!]
\plotone{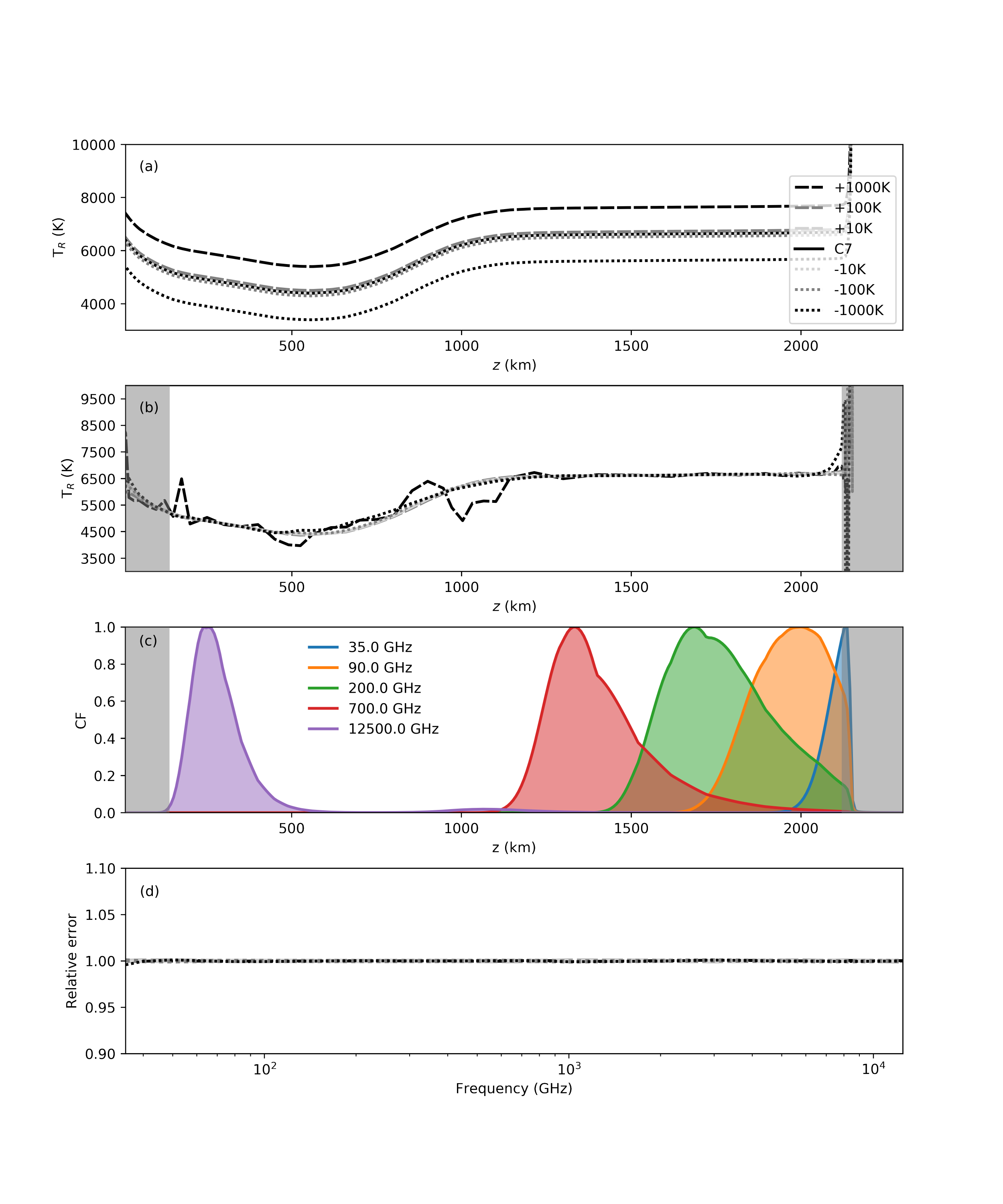}
\caption{Convergence analysis using C7 model for a set of initial conditions of radial temperature profile with $-1000$K, $-100$K, $-10$K, $+10$K, $+100$K, and $+1000$K. (a) Set of models used as initial conditions for the chromosphere using as reference the C7 model. (b) Radial temperature profile of each computed model in hydrostatic equilibrium. (c) Contribution functions of the C7 model for frequencies close to those observed by ALMA and Spitzer. (d) Relative error of the synthetic spectrum for each model taking as reference the C7 model. All the models recovery the C7 synthetic spectrum but $+1000$K model not recovery the radial temperature profile. The models $+1000$K, $-1000$K, and $+100$K shows negative values for radial temperature at altitudes higher than $2130$km over the photosphere where the model can not be applied (gray regions panel b-c).
\label{fig:tinputoutput.pdf}}
\end{figure}

We test the autoconvergence of the method using a set of variations of C7 model and found that variations greater than $+1000$K over the profile impacts in the final radial temperature model. This numerical behavior shows that the initial conditions should be lower than $+1000$K over the final solution. We can not compute directly stars with spectral class far to solar-like classification. Computations with small variations in the radial temperature profile to reproduce their emission is mandatory, specially for stars colder than the Sun. We used the radial density of C7 model as an initial condition for all the simulations. However, Figure \ref{fig:tinputoutput.pdf}c show that the CF for low frequencies is truncated due to low density at this altitude. In order to fix the CF a wider chromosphere than C7 model of density is required.  

In this context, we use spectral observations of $\alpha$ Cen A at millimeter, submillimeter, and infrared wavelengths and C7 semiempirical model as input model with the following constraints: without restrictions, limit the lower temperature, and bounding temperature profile. We found that the solution of KP is acceptable only in the regimen of "bounding temperature" as shows in the model KP2019D. A detailed simulation using greater spatial resolution shows that the shape of the profile preserves but the profile shows several spikes that are numerical artifacts
%As we mentioned, the {\bf high} spatial resolution model KP2019D produces several spikes 
in the radial temperature profile. These spikes are related to the increase in the brightness temperature in some points of the observed spectrum. Specifically at Band 3 ($97.5$GHz) and 9 ($679$GHz) of ALMA. 

We eliminate the spikes using the filter of Savitzky-Golay with a polynomial of nine degree. We compute the hydrostatic equilibrium of this last profile and called to this model KP2019E. The synthetic spectrum obtained from KP2019E fixed all the observed points inside of their observational errors for $\alpha$ Cen A except band 3 and 9 of ALMA. The spectrum computed from KP2019E suggest that both bands presents an over flux in agreement with the analysis of \citet{liseau19}.

%KP was able to fix this flux but the result included spikes in the profile that show the sensitivity of KP to flux variations over the spectrum.

\begin{figure}[ht!]
\plotone{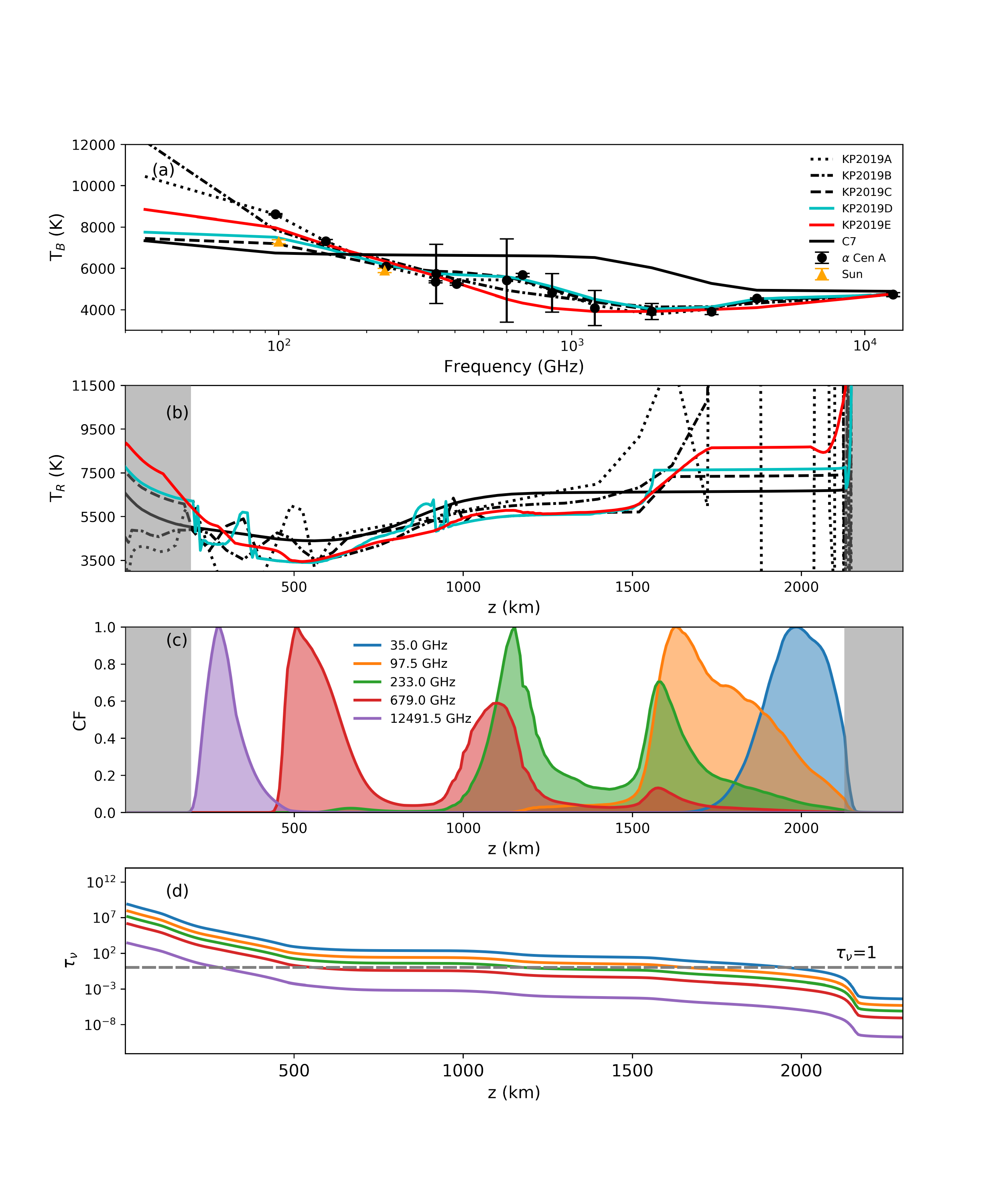}
\caption{$\alpha$ Cen A model obtained from KP. (a) The dotted line shows the model KP2019A without restrictions, dashed-point line presents the KP2019B model with positive temperatures restriction, dashed line is KP2019C model with bounded temperature, cyan solid line indicate the synthetic spectrum from the raw temperature profile (KP2019D model), the red solid line is the synthetic spectrum after smooth (KP2019E model), and the solid black line is the synthetic spectrum of C7 solar model. The circle markers are observations of $\alpha$ Cen A from  \citet{liseau16, liseau13}, and the triangle markers are the sun observations from ALMA \cite{white17}. (b) Radial temperature profiles over the photosphere. We show that the radial temperature for $\alpha$ Cen A between $290$ and $1560$km is lower than the solar model. In the outer boundary, the temperature remains higher than the solar model. (c) Normalized Contribution Function (CF) of the KP2019E model of $\alpha$ Cen A for the same frequencies. For the case of $12491$GHz, we found that the CF has a maximum close to $250$km over the photosphere. For the other cases, the CF shows three peaks for $679$GHz and two peaks for $233$GHz and one peak for $97.5$GHz and $35$GHz. (d) Optical depth for $35$GHz, $97.5$GHz, $233$GHz, $679$GHz, and $12491.5$GHz for KP2019E model of $\alpha$ Cen A. 
\label{fig:kp.pdf}
}
\end{figure}

The CF for frequencies at $35$GHz, $97.5$GHz, $233$GHz, $679$GHz, and $12491.5$GHz (ALMA and Spitzer coverage, Figure \ref{fig:kp.pdf}c) shows similarities (one peak of CF at $100$GHz and two peaks at $230$GHz) in comparison with the solar models \citep{selhorst19}. In the KP2019E model, the maximum peaks for the aforementioned frequencies occurs at a height of $2140$km, $1625$km, $1150$km, $510$km, and $275$km that corresponds with the distance over the stellar photosphere where $\tau_{\nu} \approx 1$ (Figure \ref{fig:kp.pdf}d).

The results obtained by the methodology here presented show that KP is capable of finding the physical conditions of stellar atmospheres of Solar-Like stars using their observed spectra at millimeter, submillimeter, and infrared wavelengths. We show, that KP is stable and a fast method that can compute thousands of atmospheres in a few hours. 

The impact of the radial density as initial condition in the computation of the radial temperature profile as well as the estimation of a wider solar chromosphere using our methodology is considered as future work. KP can be found under request to the authors.

\acknowledgments
The authors thank the referee for helpful and valuable
comments on this paper. This work was supported by the Ciencia B\'asica (254497) CONACyT Fellowship.

%\begin{figure}[ht!]
%\plotone{tinputoutput.pdf}
%\caption{Convergence analysis {\bf using} C7 model for a set of initial conditions of radial temperature profile {\bf with} -1000K, -100K, -10K, +10K, +100K, and +1000K. In the top panel we show the set of initial conditions {\bf models} for the chromosphere using as reference the C7 model. In the middle panel we plot the {\bf radial temperature profile of each} computed model in hydrostatic equilibrium. In the bottom panel we show the {\bf relative error of the} synthetic spectrum for each model {\bf taking as reference the C7 model. All the models except +1000K recovery the C7 synthetic spectrum. The models +1000K, -1000K, and +100K shows regions of negative values for radial temperature at altitudes higher than 2130 km over the photosphere.}
%\label{fig:tinputoutput.eps}}
%\end{figure}

%\begin{figure}[ht!]
%\plotone{kp.pdf}
%\caption{Top panel: The black solid line indicate the synthetic spectrum from the raw temperature profile {\bf (KP2019A model)}, the red line is the synthetic spectrum after smooth {\bf (KP2019B model), and the dashed line is the synthetic spectrum of C7 solar model. The circle markers} are observations {\bf of $\alpha$ Cen A} from  \citet{liseau16}{\bf, and the triangle markers are observations from ALMA \cite{white17} }.  Bottom panel: {\bf Radial temperature profiles over the photosphere used in the top panel. We show that the radial temperature for $\alpha$ Cen A between 290 and 1560 km is lower than the solar model. In the outer boundary the temperature remains higher than the solar model.}  
%\label{fig:KP.eps}
%}
%\end{figure}

%\begin{figure}[ht!]
%\plotone{tau_cf.pdf}
%\caption{{\bf Top Panel: Optical depth for 35 GHz, 97.5 GHz, 233 GHz, 679 GHz, and 12491.5 GHz for KP2019E model of $\alpha$ Cen A.
%Bottom Panel: Normalized Contribution Function (CF) of KP2019E model of $\alpha$ Cen A for the same frequencies. For the case of 12491 GHz we found that the CF have a maximum close to 250 km over the photosphere. For the other cases the CF shows three peaks for 679 GHz and two peaks for 233 GHz and one peak for 97.5 GHz and 35 GHz.} \label{fig:CF.eps}}
%\end{figure}

\begin{table*}
\centering % used for centering table 
\begin{tabular}{c c c c c c} % centered columns (3 columns) 
\hline\hline \ %inserts double horizontal lines 
   	Wavelength  & Frequency & Facility & Flux  & Brightness Temperature \\
   		$[$mm] &  $[$GHz] &  &  $[$\rm mJy] & $[$K] \\
   	\hline 
 
	3.075 & 97.5    & ALMA Band 3 & 3.33$\pm$0.01   & 8618$\pm$31  \\
	2.068 & 145     & ALMA Band 4 & 6.33$\pm$0.08   & 7316$\pm$88  \\
	1.287 & 233     & ALMA Band 6 & 13.58$\pm$0.08  & 6087$\pm$36  \\
	0.873 & 343.5   & ALMA Band 7 & 26.06$\pm$0.19  & 5351$\pm$49  \\
	0.740 & 405     & ALMA Band 8 & 35.32$\pm$0.21  & 5242$\pm$55  \\
	0.442 & 679     & ALMA Band 9 & 107.20$\pm$1.50 & 5678$\pm$79  \\
	
\hline	
    
\end{tabular}
\caption{ALMA Brightness temperatures for $\alpha$ Cen A from \citet{liseau16}.}
\label{tab:almaobs}%ss:syntheticspectrum
\end{table*}

\begin{table*}

\centering % used for centering table 
\begin{tabular}{c c c c c c} % centered columns (3 columns) 
\hline\hline \ %inserts double horizontal lines 
   	Wavelength  & Frequency & Facility & Flux  & Brightness Temperature \\
   		$[$mm] &  $[$GHz] &  &  [$\rm mJy$] & [K] \\
   	\hline 
 
	0.870 & 345     & LABOCA               & 28.00$\pm$7     & 5738$\pm$1432 \\
	0.500 & 600     & Herschel-SPIRE       & 80$\pm$30       & 5421$\pm$2018 \\
	0.350 & 857     & Herschel-SPIRE       & 145.00$\pm$28   & 4822$\pm$927  \\
	0.250 & 1199    & Herschel-SPIRE       & 240.00$\pm$50   & 4084$\pm$845  \\
	0.160 & 1874    & Herschel-PACS        & 560.00$\pm$60   & 3920$\pm$394  \\
	0.100 & 2998    & Herschel-PACS        & 1410.00$\pm$50  & 3909$\pm$135  \\
	0.070 & 4283    & Herschel-PACS        & 3350$\pm$28     & 4540$\pm$37   \\
	0.024 & 12491.5 & Spitzer-MIPS         & 28530$\pm$580   & 47364$\pm$91   \\
	
\hline	
    
\end{tabular}
\caption{ Brightness temperatures at infrared wavelengths for $\alpha$ Cen A from \citet{liseau13}.}
\label{tab:almaobs2}%ss:syntheticspectrum
\end{table*}

\begin{table*}

\centering % used for centering table 
\begin{tabular}{c  c  c  c} % centered columns (3 columns) 
\hline\hline \ %inserts double horizontal lines 
  Frequency  & Brightness Temperature & Frequency  & Brightness Temperature \\
  $[$GHz]  &  [K] & $[$GHz]  &  [K] \\
   	\hline 
    
    35	&	7298 	&	800	&	6600 \\
    40	&	7150 	&	900	&	6591 \\
    50	&	6975 	&	1000	&	6577 \\
    60	&	6878 	&	2000	&	5918 \\
    70	&	6820 	&	3000	&	5268 \\
    80	&	6781 	&	4000	&	4982 \\
    90	&	6754 	&	5000	&	4866 \\
    100	&	6734 	&	6000	&	4823 \\
    200	&	6662 	&	7000	&	4831 \\
    300	&	6641 	&	8000	&	4838 \\
    400	&	6630 	&	9000	&	4844 \\
    500	&	6621 	&	10000	&	4854 \\
    600	&	6615 	&	11000   &	4866 \\
    700	&	6608 	&	12500	&	4886 \\  	

%   	35	&	7299	&	772	    &	6603 \\	
%    48	&	7002	&	1052	&	6568 \\	
%    65	&	6846	&	1433	&	6397 \\	
%    89	&	6757	&	1953	&	5959 \\	
%   121	&	6707	&	2661	&	5438 \\	
%    164	&	6677	&	3626	&	5062 \\	
%    224	&	6656	&	4941	&	4871 \\	
%   305	&	6641	&	6733	&	4823 \\	
%    416	&	6629	&	9174	&	4846 \\	
%    567	&	6617	&	12500	&	4887 \\
   	
%        35	&	7299 \\
%        48	&	7002 \\
%        65	&	6846 \\
%        89	&	6757 \\
%       121	&	6707 \\
%        164	&	6677 \\
%        224	&	6656 \\
%        305	&	6641 \\
%        416	&	6629 \\
%        567	&	6617 \\
%    772	    &	6603 \\
%    1052	&	6568 \\
%   1433	&	6397 \\
%   1953	&	5959 \\
%    2661	&	5438 \\
%    3626	&	5062 \\
%    4941	&	4871 \\
%    6733	&	4823 \\
%    9174	&	4846 \\
%    12500	&	4887 \\
\hline	
    
\end{tabular}
\caption{Synthetic spectrum between 35 GHz and 12.5 THz by PAKALMPI using the C7 model. }
\label{tab:ss}%ss:syntheticspectrum
\end{table*}



\end{document}